\documentstyle[prd,floats,aps]{revtex}
\input psfig
\begin{document}
\title{Development of a Double Pendulum for Gravitational Wave Detectors}

\author{ Mark A. Beilby, Gabriela Gonzalez, Michelle Duffy, Amber
Stuver, Jennifer Poker.}

\address{Department of Physics, Pennsylvania State University, University Park, PA 16802.}
\maketitle
\begin{abstract}
Seismic noise will be the dominant source of noise at low frequencies
for ground based gravitational wave detectors, such as LIGO now under
construction. Future interferometers installed at LIGO plan to use at
least a double pendulum suspension for the test masses to help filter
the seismic noise.  We are constructing an apparatus to use as a test
bed for double pendulum design.  Some of the tests we plan to conduct
include: dynamic ranges of actuators, and how to split control between
the intermediate mass and lower test mass; measurements of seismic
transfer functions; measurements of actuator and mechanical cross
couplings; and measurements of the noise from sensors and actuators.
All these properties will be studied as a function of mechanical
design of the double pendulum.
\end{abstract}
\vspace{-6.5cm} 
\begin{flushright}
\baselineskip=15pt
CGPG-99/11-5  \\
gr-qc/9911027\\
\end{flushright}
\vspace{5.5cm}

The next upgrade installed at LIGO \cite{Ab} plans to use at least a double
pendulum suspension for the test masses to help filter seismic noise.
We are constructing a facility to be used as a test-bed for testing
the mechanical and local control design of test mass suspension
systems for interferometer gravitational wave detectors.  The basic
design that we are following is based on the GEO gravitational wave
detector suspension design, which uses multiple pendulums \cite{Pl}.  Our
prototype is shown schematically in Figure 1. 
\begin{figure}
\centerline{\psfig{file=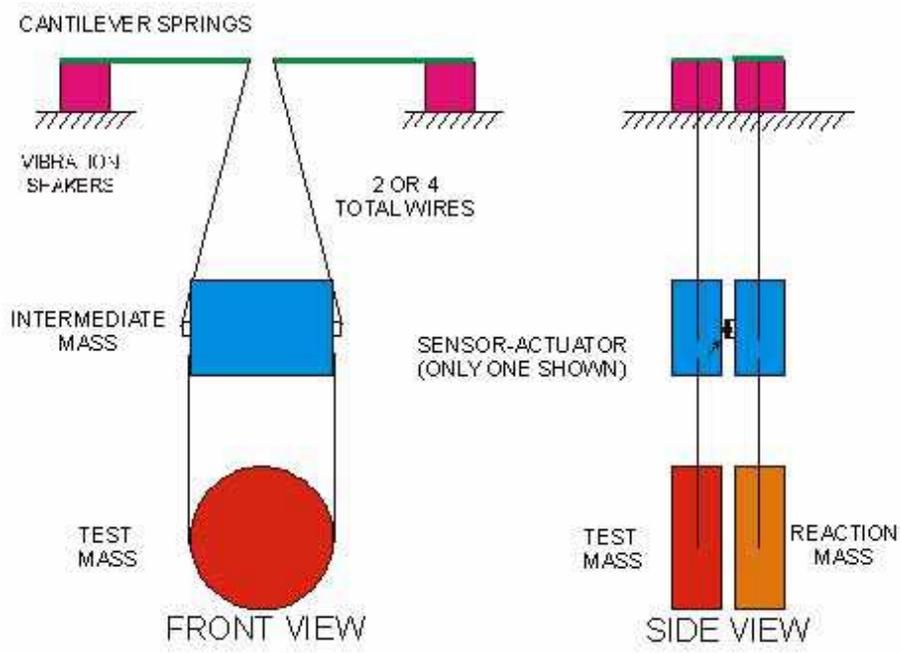,height=9cm}}
\centerline{\caption{Prototype Double-Double Pendulum.}}
\end{figure}
In addition to the
double pendulum, which includes the test mass, a second identical
double pendulum is just behind the test mass pendulum.  This second
pendulum acts as a reaction pendulum, which is used to hold the
sensors and actuators used to control the test position of the test
mass.  Putting the sensors and actuators on an identical second
pendulums allows a smaller relative motion between the actuators and
the test mass.  In addition to gaining extra seismic filtering of a
double pendulum, the double pendulum also allows the possibility of
controlling the test mass at the intermediate mass rather than from
the test mass itself.  Therefore, the magnets used for position
control, which unfortunately ruin the high mechanical Q of the test
mass, can be moved to the intermediate mass.  The disadvantage of a
double pendulum is that it is more complex to model and hence control,
so that testing of actual configurations is a critical activity. The
double pendulum is hung from cantilever spring, so that there is also
vertical seismic isolation.  One feature of our test bed facility is
that the design is made flexible so that a variety of test mass
configurations can be tested, such as changing the positions of the
sensors and actuators. Another feature in our test bed facility is the
use of a high precision three-axis vibration shaker used for
diagnostics.  The construction and assembly of the mechanical pieces
of the initial single intermediate pendulum supported by cantilever
springs is near completion.  The first goal is to control this single
pendulum in all six degrees of freedoms.  Testing and calibration of
position controller: sensors (LED shadow detectors) and actuators
(magnets and coils) are now in progress.  Testing and characterizing
the three-axis vibration shaker has been in progress.  The next phase
will be to build a double pendulum with a dummy test mass of the size
of the test masses currently being installed in LIGO.  Finally, a
double pendulum, with its corresponding double reaction mass pendulum
will be built and tested.  Some of the tests we plan to conduct
include: dynamic ranges of actuators, used to control the position of
the double pendulum masses and how to split control between the
intermediate mass and lower test mass; measurements of seismic
transfer functions of the double pendulum; measurements of actuator
and mechanical cross couplings; and measurements of the noise from
sensors and actuators.  All these properties will be studied as a
function of mechanical design of the double pendulum, such as two
versus four wire suspension, wire attachment points (which determine
the resonant frequencies of the pendulum), actuator and sensor
placement, intermediate mass shape and size, cantilever spring design
and number, and modal damping versus point to point damping.

\acknowledgements

We would like to thank Dr. Mike Plissi of the GEO Project for his interest in this work. This research is supported by The Pennsylvania State University and NSF Grant No. PHY-9870032.

\end{document}